\documentclass{article}
\usepackage{graphicx}
\usepackage{epsfig}
\usepackage{amsfonts}
\usepackage{cite}

\usepackage{a4}

\usepackage{amsmath}

\begin{document}
\title
{\bf Embedding Gauss-Bonnet scalarization models   \\
in higher dimensional topological theories 
}

\author{
{\large Carlos Herdeiro}$^{\dagger}$,
{\large Eugen Radu}$^{\dagger}$ 
{\large}and {\large D. H. Tchrakian}$^{\ddagger \star}$   
\\  
\\
$^{\dagger}$ {\small Departamento de Matem\'atica da Universidade de Aveiro and } 
\\
 {\small  Center for Research and Development  in Mathematics and Applications (CIDMA),}   
\\
   {\small Campus de Santiago, 3810-183 Aveiro, Portugal}
	\\ 
$^{\ddagger}${\small School of Theoretical Physics, Dublin Institute for Advanced Studies,
10 Burlington Road, Dublin 4, Ireland}\\
$^{\star}${{\small Department of Computer Science, NUIM, Maynooth, Ireland}}
}
\date{}
\newcommand{\dd}{\mbox{d}}
\newcommand{\tr}{\mbox{tr}}
\newcommand{\la}{\lambda}
\newcommand{\bt}{\beta}
\newcommand{\del}{\delta}
\newcommand{\ep}{\epsilon}
\newcommand{\ta}{\theta}
\newcommand{\ka}{\kappa}
\newcommand{\f}{\phi}
\newcommand{\vf}{\varphi}
\newcommand{\F}{\Phi}
\newcommand{\al}{\alpha}
\newcommand{\ga}{\gamma}
\newcommand{\de}{\delta}
\newcommand{\si}{\sigma}
\newcommand{\Si}{\Sigma}
\newcommand{\bnabla}{\mbox{\boldmath $\nabla$}}
\newcommand{\bomega}{\mbox{\boldmath $\omega$}}
\newcommand{\bOmega}{\mbox{\boldmath $\Omega$}}
\newcommand{\bsi}{\mbox{\boldmath $\sigma$}}
\newcommand{\bchi}{\mbox{\boldmath $\chi$}}
\newcommand{\bal}{\mbox{\boldmath $\alpha$}}
\newcommand{\bpsi}{\mbox{\boldmath $\psi$}}
\newcommand{\brho}{\mbox{\boldmath $\varrho$}}
\newcommand{\beps}{\mbox{\boldmath $\varepsilon$}}
\newcommand{\bxi}{\mbox{\boldmath $\xi$}}
\newcommand{\bbeta}{\mbox{\boldmath $\beta$}}
\newcommand{\ee}{\end{equation}}
\newcommand{\eea}{\end{eqnarray}}
\newcommand{\be}{\begin{equation}}
\newcommand{\bea}{\begin{eqnarray}}

\newcommand{\z}{\zeta}

\newcommand{\ii}{\mbox{i}}
\newcommand{\e}{\mbox{e}}
\newcommand{\pa}{\partial}
\newcommand{\Om}{\Omega}
\newcommand{\om}{\omega}
\newcommand{\vep}{\varepsilon}
\newcommand{\bfph}{{\bf \phi}}
\newcommand{\lm}{\lambda}
\def\theequation{\arabic{equation}}
\renewcommand{\thefootnote}{\fnsymbol{footnote}}
\newcommand{\re}[1]{(\ref{#1})}
\newcommand{\R}{{\rm I \hspace{-0.52ex} R}}
\newcommand{\N}{{\sf N\hspace*{-1.0ex}\rule{0.15ex}%
{1.3ex}\hspace*{1.0ex}}}
\newcommand{\Q}{{\sf Q\hspace*{-1.1ex}\rule{0.15ex}%
{1.5ex}\hspace*{1.1ex}}}
\newcommand{\C}{{\sf C\hspace*{-0.9ex}\rule{0.15ex}%
{1.3ex}\hspace*{0.9ex}}}
\newcommand{\eins}{1\hspace{-0.56ex}{\rm I}}
\renewcommand{\thefootnote}{\arabic{footnote}}

\maketitle


\bigskip

\begin{abstract}
In the presence of appropriate non-minimal couplings between a scalar field and the curvature squared Gauss-Bonnet (GB) term, 
compact objects such as neutron stars and black  holes (BHs) can spontaneously scalarize, becoming a preferred vacuum. Such strong gravity
phase transitions have attracted considerable attention recently. The non-minimal coupling functions that allow this mechanism are, however,  
always postulated \textit{ad hoc}. Here we point out that families of such functions naturally emerge in the context of Higgs--Chern-Simons 
gravity models, which are found as dimensionally descents of higher dimensional, purely topological,  Chern-Pontryagin non-Abelian densities.
As a proof of concept, we study spherically symmetric scalarized BH solutions in a particular Einstein-GB-scalar field model, whose coupling
is obtained from this construction, pointing out novel features and caveats thereof.
The possibility of vectorization is also discussed, since this construction also originates vector fields  non-minimally coupled to the GB  invariant.

\end{abstract}

\section{Introduction  }

The subject of 'spontaneous scalarization' of asymptotically flat black holes (BHs) has received considerable interest over the last several years.
This phenomenon occurs due to non-minimal couplings  in the scalar field action,
 which
allow circumventing well-known no-hair theorems \cite{Herdeiro:2015waa}.

The typical non-minimal coupling is between a real
scalar field $\phi$ and some source term ${\cal I}$; it triggers a repulsive gravitational effect, via an effective
tachyonic mass for $\phi$. 
As a result, the General Relativity 
(GR) 
BH solutions are unstable against scalar perturbations in regions
where the source term is significant, leading  to  BH scalar hair growth.  

Following \cite{Astefanesei:2020qxk}, let us briefly review this mechanism, restricting to $D=4$
spacetime dimensions.
The starting point here is the action for the scalar sector,
which has a generic form
\begin{eqnarray}
\label{action0}
\mathcal{S}_\phi= \int d^4 x \sqrt{-g} 
\left[
 \frac{1}{2}(\nabla \phi)^2 
+\alpha f(\phi) {\cal I}(\psi;g)
\right] \ ,
\end{eqnarray}
with $f(\phi) $ the
{\it coupling function},
 $\alpha$  a  coupling constant,
while
the  source term ${\cal I}$ generically depends on 
some extra-matter field(s)
$\psi$
and on   the
metric tensor $g_{\mu\nu}$.
The corresponding equation of motion for the scalar field $\phi$
reads
\begin{eqnarray}
\label{eq-phi}
\nabla^2\phi= \alpha \frac{d f}{d \phi}{\cal I} \ .
\end{eqnarray}
An essential feature of a model allowing for scalarization 
is the existence of a 
 fundamental solution of the above equation,
\begin{eqnarray}
\label{c0}
 \phi=\phi_0,~~~{\rm with}~~~\frac{d f}{d \phi}\Big |_{\phi=\phi_0}=0 \ ,
\end{eqnarray}
providing the 
 {\it `ground state'} of the scalar model.\footnote{`Ground state' is intended to mean an equilibrium solution, not necessarily stable.}
As a result, the usual GR solutions (with $\phi=\phi_0$) solve also the considered model 
(which consists in 
(\ref{action0}) supplemented with terms for gravity and matter field(s) $\psi$),  being the {\it fundamental solutions} of the model.

Apart from the  ground state, the model possesses a second set of solutions, with a nontrivial scalar field -- {\it the scalarized BHs}.
Moreover, usually\footnote{There are exceptions for special coupling functions; in some models, 
 the scalarized BHs do not emerge as an instability  of the fundamental solutions \cite{Blazquez-Salcedo:2020nhs}.} 
 they are smoothly connected with the fundamental set, which is approached for $\phi= \phi_0$.
Then,
at the linear level, spontaneous scalarization manifests
itself as a tachyonic instability triggered by a negative
{\it effective} mass squared of the scalar field.

Around the ground state, 
the coupling function possesses the following expression
(with $\delta \phi=\phi-\phi_0$)
\begin{eqnarray}
\label{f-phi-small}
f(\phi)=f|_{\phi=\phi_0}+ \frac{1}{2} \frac{d^2 f}{d \phi^2}\Big |_{\phi=\phi_0}  \delta \phi^2+\dots \ .
\end{eqnarray}
Then the linearized form of equation (\ref{eq-phi})
($i.e.$ with a small-$\phi$) is:
\begin{eqnarray}
\label{eq-phi-small}
(\nabla^2-\mu_{\rm eff}^2)\delta \phi =0,~~{\rm where}~~ \mu_{\rm eff}^2= 
\alpha \frac{d^2 f}{d \phi^2}\Big |_{\phi= \phi_0} {\cal I} .
\end{eqnarray}
There are time independent solutions (bound states) of the above equation describing {\it scalar
clouds}: for appropriate choices of   ${\cal I}$\,  
the tachyonic condition ($\mu_{\rm  eff}^2<0$) can be satisfied 
 for a specific (discrete) set of backgrounds,
as specified $e.g.$
by their global charge(s). 
The onset of the instability is marked by such bound states. 
The backreacting continuation of the scalar clouds result in scalarized BHs.

Various choices of the source term ${\cal I}$ have been considered in the literature.
In the context of this work, of special interest is the case of geometric scalarization, with
\begin{eqnarray}
\label{LGB}
{\cal I}=R^{\mu\nu\rho\sigma}R_{\mu\nu\rho\sigma}-4~R^{\mu\nu}R_{\mu\nu}+R^2=L_{GB}~,
\end{eqnarray}
the Gauss-Bonnet (GB) invariant,
a choice which allows for the scalarization of vacuum Schwarzschild BH (we recall that 
$L_{GB}=-48 M^2/r^6$
in this case, with $M$ the BH mass). 
This model has been extensively studied, starting with
\cite{Silva:2017uqg,Doneva:2017bvd,Antoniou:2017acq,Antoniou:2017hxj}
where the first examples of  scalarized BHs resulting from this type of mechanism have been 
reported. 
Further work includes the study of scalarized BHs in various extensions
of the initial framework
\cite{Minamitsuji:2018xde,
Brihaye:2018grv,
Macedo:2019sem,
Doneva:2019vuh,
Andreou:2019ikc,
Minamitsuji:2019iwp,
Blazquez-Salcedo:2020crd,
Guo:2020zqm,
Bakopoulos:2020dfg,
Peng:2020znl,
Liu:2020yqa,
Cardoso:2020cwo,
Ventagli:2020rnx,
Guo:2020sdu,
Doneva:2020qww,
Heydari-Fard:2020iiu,
Bakopoulos:2020mmy}
and
the investigation of solutions' stability
\cite{Blazquez-Salcedo:2018jnn,
Silva:2018qhn,
Blazquez-Salcedo:2020rhf,
Blazquez-Salcedo:2020caw};
furthermore,
partial analytical results are reported in Refs.
\cite{Hod:2019pmb,
Hod:2019vut,
Konoplya:2019goy,
Hod:2020jjy},
while scalarized, rotating  BHs 
are studied in 
 \cite{Cunha:2019dwb,
Collodel:2019kkx,
Doneva:2021dqn,
Dima:2020yac,
Herdeiro:2020wei,
Berti:2020kgk,
Doneva:2020nbb}.

The explicit form of the coupling function
does not appear for be important for the existence of scalarized solutions (although it impacts on their properties),
as long as the conditions discussed above
are satisfied.
For example, the results in \cite{Silva:2017uqg}
are found for 
$f(\phi)=  \phi^2$,
while Ref. 
\cite{Doneva:2017bvd},
considers a coupling function
$f(\phi)= 1-e^{-6\phi^2}$.

\medskip

To the  best of our knowledge, 
a common feature of  models 
allowing for BH scalarization is that
the origin of the term   $f(\phi)  L_{GB}$ in (\ref{action0}), and in particular
the choice of the coupling function $f(\phi) $,
is 
{\it `ad hoc'}, 
missing a well motivated origin\footnote{
The coupling function $f(\phi)=e^{- \phi}$ naturally
appears in the string theory context, when including first-order
$ \alpha'$ corrections
(with $\phi$ the dilaton field). This choice, however, does not allow for 
BH scalarization, the condition (\ref{c0}) being not satisfied for any finite $\phi_0$.
Despite this fact, some features found for scalarized BHs
occur also in this case, a relevant example being 
the appearance of repulsive effects for static, spherically
 symmetric solutions 
 \cite{Buonanno:1997mi}.
}.
The main purpose of this work is the study the 
basic properties of the BH solutions in a model where
the intraction term $f(\phi)  L_{GB}$  
emerges naturally from a Higgs-Chern-Simons gravity (HCSG), 
originally proposed in \cite{Tchrakian:2017fdw,Radu:2020ytf}.
The corresponding expression of 
 the coupling function  is
\begin{eqnarray}
\label{f}
f(\phi)= \phi \left(1-\frac{1}{3}\phi^2\right)~.
\end{eqnarray}
The   field $\phi$ here
is a Higgs-like scalar field,
being a relic of
a Yang-Mills (YM) connection in  higher dimensions,
and
approaches a nonzero value at infinity,
with two discontinuous vacua at $\phi_0=\pm  1$.
As we shall see, the expression (\ref{f}) 
of the coupling function
allows for scalarization of Schwarzschild BHs; 
 the basic properties of the 
scalarized
solutions  are rather similar to those in  Ref. \cite{Silva:2017uqg}
(where a quadratic coupling function has been employed).
However,
there are also  novel features;
 an interesting one is that  
scalarization occurs for both signs of the constant $\alpha$ 
(which is not the case for other models with scalarized static BHs).
Another interesting feature is the existence 
of an extension of the model with a vector field, which allows for
vectorization of the generic Einstein-GB-scalar (EGBs) BHs.

\medskip

This paper is organized as follows. 
In Section  2 we review the basic features of
the model in \cite{Tchrakian:2017fdw}, in particular its derivation
starting with a Chern-Pontryagin (CP) density in $D=8$ dimensions.
The solutions of the model are reported in 
Section 3.
Both generic  configurations 
(with a value of the scalar field 
which does not approach asymptotically the ground state)
and scalarized BHs are discussed; moreover, a perturbative (analytic) 
solution is derived in the former case. 
Working in the probe limit,
the solutions of an extension of the original model with an extra-vector field
are also reported.
We conclude with a summary and a discussion in Section 4.

\section{HCS gravity }

The HCSG models in 
\cite{Tchrakian:2017fdw,Radu:2020ytf}
are particular examples of gauge theories of gravity,
and follow the spirit (and the  general framework) in 
\cite{Utiyama:1956sy,Kibble:1961ba,Witten:1988hc,Chamseddine:1989nu,Chamseddine:1990gk,Deser:1981wh},
with
the usual identification of the spin-connection and the Riemann curvature with the YM connection and curvature.
As a new feature, an extra-Higgs-like scalar and a vector field occur  in the four dimensional 
action for the specific model studied in this work.
For completeness, in this section 
we briefly review the flavour of the results in Ref. \cite{Tchrakian:2017fdw,Radu:2020ytf},
 which proposes 
 a general framework for the construction of such HCSG models.

The starting point in this construction is the
CP density for a 
$SO(2n)$ YM field in $D=2n$ dimensions ($n=2,3,\dots$),
\begin{eqnarray}
\label{CP}
\Om_{\rm CP}= \mbox{Tr}\, F\wedge F\wedge \dots\wedge F ~~~ ( n\ {\rm times})~,
\end{eqnarray}
with $F$ the curvature 2-form.
By construction, 
the CP density
can be expressed locally as a total divergence,
$\Om_{\rm CP} = \nabla_M \bar \Omega^M$
($M=1,2,\dots,D$).

In the usual approach, a  Chern-Simons density is defined as the $D^{\rm th}$ component of 
$ \bar \Omega^M$,
which results in a YM theory in a $d=2n-1$ odd-dimensional spacetime\footnote{
At no point is a metric involved in this construction; this is a topological theory.}.
However, 
the approach in \cite{Tchrakian:2017fdw,Radu:2020ytf}
(see also the 
 Refs.~\cite{Tchrakian:2010ar,Radu:2011zy} 
and the Appendix A of Ref.~\cite{Tchrakian:2015pka})
introduces an extra-step, by considering first the dimensional descent of the CP density (\ref{CP}) 
to some intermediate dimension $d<D=2n$,
which can be {\it odd} or {\it even}.
As usual with gauge fields, 
the relics of the gauge connection on the co-dimension(s) are Higgs scalar(s).
The remarkable property of the resulting density (dubbed now
Higgs-CP density $\Om_{\rm HCP}$, being given in terms of both the residual gauge field and the Higgs scalar $\F$), 
is that, like the original CP density, it
is also a $total\ divergence$,
\begin{eqnarray}
\label{W}
\Om_{\rm HCP}=\nabla_i   W^i\ ,\quad i=1,.., d.
\end{eqnarray}
As with the definition of the usual Chern-Simons densities,
the resulting 
 HCS density in $d -1$ dimensions is defined
as the $d^{\rm th}$ component of $ W^i$.

For both the Chern-Simons and Higgs-Chern-Simons cases, the final step is the  (standard) prescription 
for the passage to gravity, 
 the spin-connection being identified with the YM connection.
In the  latter case, 
this prescription  must be extended by
the corresponding elements pertaining to the Higgs sector, which generically result in 
extra frame-vector fields, apart from the scalar field(s)  \cite{Tchrakian:2017fdw,Radu:2020ytf}.
By analogy to the standard Chern-Simons gravities (which exist in odd-dimensions only \cite{Tchrakian:2019iem}),
the resulting models are dubbed HCSG.

\medskip

Let us exemplify
the generic construction
 with the simplest two cases, the starting point being the  CP densities
(\ref{CP})
 in $D=6,8$ dimensions.
The case of interest here is $d=5$, the final Higgs-Chern-Simons being defined in four dimensions,
with the following Lagrangians:
\bea
\Omega^{(4,6)}_{\rm HCS}&=&\vep^{\mu\nu\rho\si}\,\mbox{Tr}\ F_{\mu\nu}\,F_{\rho\si}\,\F~,
\label{HCS46}
\\                  
\Omega^{(4,8)}_{\rm HCS}&=&\vep^{\mu\nu\rho\si}\,\mbox{Tr}\bigg[
\F\left(  F_{\mu\nu}F_{\rho\si}+\frac29\,\F^2\,F_{\mu\nu}F_{\rho\si}+\frac19\,F_{\mu\nu}\F^2F_{\rho\si}\right)
\nonumber
\\
&&\qquad\qquad\qquad-\frac29
\left(\F D_{\mu}\F D_{\nu}\F-D_{\mu}\F\F D_{\nu}\F+D_{\mu}\F D_{\nu}\F\F\right)F_{\rho\si}\bigg]~,
\label{HCS48}
\eea
these being the $5^{\rm th}$ components of the corresponding $W^i$ vector in (\ref{W}).
Also, note that the $d=5$ HCP density leading to $\Omega^{(4,8)}$
 is found by considering the reduction
of the $D=8$ CP density over a three dimensional sphere of unit radius.
Also, the gauge group in \re{HCS46}-\re{HCS48} is chosen to be $SO(5)$ while the Higgs
field takes it values in the orthogonal complement of $SO(5)$ in $SO(6)$.
 
After 
the passage to gravity,
the corresponding  HCSG Lagrangians
read
\cite{Tchrakian:2017fdw}
\begin{eqnarray}
\label{HCSG46}
{\cal L}^{(4,6)} 
&=&
 \vep^{\mu\nu\rho\si}\vep_{abcd}\,\f\,{R}_{\mu\nu}^{ab}{R}_{\rho\si}^{cd}~,
\label{f46}
\\
\label{HCSG48}
{\cal L}^{(4,8)}
&=&
 \vep^{\mu\nu\rho\si}\vep_{abcd}\left\{\left(1
-\frac13\left(\phi^2+A^2\right)\right)\f\,{R}_{\mu\nu}^{ab}{R}_{\rho\si}^{cd}
+\frac83\,{R}_{\mu\nu}^{ab}\,
\nabla_{\rho}A^{c}(\f \nabla_{\si}A^{d}-2 A^{d} \f_{,\si})\right\}~.
\end{eqnarray}
The  scalar $\f$  and the `frame-vector field' $A_a$ are relics of the Higgs scalar 
(with $A^2=A_\mu A^\mu)$.
The density (\ref{HCSG48}) can be cast in a more useful form by dropping a total derivative term,
 which results
in the equivalent expression\footnote{Note that
$
\vep^{\mu\nu\rho\si}\vep_{abcd}  
{R}_{\mu\nu}^{ab}{R}_{\rho\si}^{cd} =4 L_{GB}.
$
}
\begin{eqnarray}
\label{Ls}   
{\cal L}^{(4,8)} = 
\vep^{\mu\nu\rho\si}\vep_{abcd}\,\f\,\left\{\left(
1-\frac13\f^2-A^2\right){R}_{\mu\nu}^{ab}{R}_{\rho\si}^{cd}
+8\,R_{\mu\nu}^{ab} \nabla_\rho A^c \nabla_\sigma A^d
\right\}  .
\end{eqnarray}
Note that a similar construction can also be carried out 
starting with a CP density in 
 $D=2n>8$.
This results, however, in much more complicated expressions
of the corresponding HCSG Lagrangians.


\section{The BH solutions}
The Lagrangian of the full model consists in the 
usual Einstein term for gravity and kinetic term for the scalar 
field,
together with the
interaction term (\ref{HCSG46}) or 
(\ref{Ls}).

The solutions of the model with an 
interaction term (\ref{HCSG46}) 
($i.e.$
with a linear coupling function $f(\phi)=\phi$)
have been extensively studied in the literature\footnote{For a review of the literature
together with an investigation of spinning solutions, see the recent work \cite{Delgado:2020rev}.},
 starting with 
Refs. \cite{Sotiriou:2013qea,Sotiriou:2014pfa},
falling within
the Horndeski class of scalar-tensor theories of gravity \cite{Horndeski:1974wa}.
The results in this work show that 
the term $\phi L_{GB}$
naturally occurs also in this topological construction leading  to HCSG models.
This model is rather special,  
the equations of motion
being invariant
under the transformation
$\phi \to \phi+\beta$
(with $\beta$ arbitrary),
which leads to a conserved current.
The coupling function  $f(\phi)=\phi$, however, does not allow for
BH scalarization:
the condition
(\ref{c0}) is not fulfilled.

\medskip

In this work we shall focus on the interaction term 
(\ref{Ls}),  $i.e.$ with a cubic coupling function as given by (\ref{f}).
As such,  
	the full action of the model reads  
\begin{eqnarray}  
\label{action}
S=
\int d^4x \sqrt{-g} \left[  R - 2(\partial_a \phi)^2-F^2
 -\alpha {\cal L}^{(4,8)}
\right ]
, 
\end{eqnarray} 
where $F^2=F_{\mu \nu}F^{\mu \nu}$ is the standard kinetic term for a vector field
(with $F_{\mu \nu}=\nabla _\mu A_\nu-\nabla_\nu A_\mu)$.
Let us remark that the vector field $A$ does not possess the $U(1)$ gauge invariance of a Maxwell model. 

The corresponding equations of motion are found by 
varying the action~(\ref{action}) with respect to the metric tensor
 $g_{\mu \nu}$,  scalar field $\phi$
and vector field $A_\mu$.
One can easily verify that 
$A=0$ 
solves the  equations
for the vector field.
As such, to simplify the study, we
shall restrict our study mainly to the scalar-tensor case.
Some comments on the general case with $A\neq 0$
can be found
at the end of this Section.

Setting $A=0$, the scalar field possess two ground states at $\phi \equiv \phi_0=\pm 1$.
Also, let us remark that the model is invariant under the transformation
\begin{eqnarray}  
\label{tr1}
 \phi \to -\phi,~~\alpha \to -\alpha~; 
\end{eqnarray} 
we shall restrict our study to the ground state $\phi_0=1$, only. 
An important physical consequence of
the symmetry
 (\ref{tr1})
is that, in contrast to other models in the literature 
\cite{Silva:2017uqg,Doneva:2017bvd},
 the scalarization of 
Schwarzschild BH occurs for both  signs of 
the coupling constant
$\alpha$. 

\subsection{Einstein-GB-scalar field BHs}

Restricting our study to spherically symmetric solutions,  
 we consider an Ansatz with
\begin{eqnarray}
\label{ansatz}
ds^2=-e^{-2\delta(r)}N(r)dt^2+\frac{dr^2}{N(r)}+r^2(d\theta^2+\sin^2\theta d\varphi^2)  \ ,
~~{\rm with}~~N(r)=1-\frac{2m(r)}{r}~~~
{\rm and}~~ 
~\phi \equiv \phi(r) \ .
\end{eqnarray}
This ansatz results 
in two first order equations
for the functions $m,\delta$
 and a second order equation  for  $\phi$. There is also an extra second order constraint equation, which is used in practice 
to monitor the accuracy of the numerical results.
 
In this work we shall consider non-extremal\footnote{The considered EGBs model is unlikely to possess
regular (on and outside a horizon) extremal BH solutions.
A clear indication in this direction is the absence of 
a generalization  of the Bertotti-Robinson solution, 
with a metric $AdS_2\times S^2$
(that it, no  
 attractor solutions exists here).
 However, the situation may change for the  model  
with an extra vector field.
} 
BHs only,   
with a horizon located at $r=r_H>0$. 
At the horizon, the solution possesses  a power series expansion in $r-r_H$, 
that depends only on the  parameters
 $r_H$,
$\phi(r_H)$,
 $\delta(r_H)$,
and $\alpha$, the first terms being
\begin{eqnarray}
\label{eh}
m(r) =\frac{r_H}{2}+ m_1(r-r_H) +  \cdots \ , \quad
			\delta (r)  = \delta _H +\delta _1  (r-r_H)+\cdots\ ,
\quad 
\phi ( r) =  \phi _H + \phi_1 (r-r_H)+\cdots\ . 
\end{eqnarray}
The coefficient 
$\phi'(r_H)=\phi_1$
satisfies a  second order algebraic equation of the form
\begin{eqnarray}
\label{eq2}
\phi_1^2+p \phi_1+q=0 \ ,
\end{eqnarray}
where $(p,q)$ are non-trivial functions of $r_H,\phi_H$ and $\alpha$.
Consequently,  a real solution of~(\ref{eq2}) exists only if $ p^2-4q \geq  0$,
a condition which translates into the following inequality 
\begin{eqnarray} 
\label{cond}
\Delta \equiv 1-\frac{384 \alpha^2 }{r_H^4}(1-\phi_H^2)^2 \geq 0 ~,
\end{eqnarray}
which implies the existence of a minimal horizon size, 
determined by $\alpha$
and the value of the scalar field at the horizon.
As with other EGBs models,
a possible interpretation  of (\ref{cond})
is that the GB term provides a repulsive
contribution, 
which
becomes overwhelming for sufficiently small BHs, and thus prevent the existence
of an event horizon.

The approximate form of the solutions in the far field reads
\begin{equation}
\label{inf}
m(r)=N-\frac{Q_s^2}{2r }-\frac{QM _s^2}{2r^2 }+\dots\ , \quad \delta(r)= \frac{Q_s^2}{2r^2}+\dots\ ,\quad
 \phi(r)=\phi_\infty+\frac{Q_s}{r}+\frac{MQ_s}{r^2}+\dots\ , \quad 
\end{equation} 
the essential parameters here being
$M$ (the ADM mass), $\phi_\infty $
and
$Q_s$
(the scalar `charge').

Other physical quantities of interest are found from the horizon data.
These are
the Hawking temperature, $T_H$, 
and the event horizon area, $A_H$,  
\begin{eqnarray}
T_H=\frac{1}{4\pi} N'(r_H)e^{-\delta(r_h)} \ , \qquad A_H=4\pi r_H^2 \ ,
\end{eqnarray}
together with the 
  BH entropy, which is the sum of two terms\footnote{The expression of the
coupling function
$f(\phi)$ in (\ref{S}) is 
\begin{eqnarray}
\label{fn}
 f(\phi)=\phi\left(1-\frac{\phi^2}{3}\right)-\frac{2}{3}~,
\end{eqnarray}
which guarantees that a
scalar field in the ground state $\phi \equiv 1$
provides no contribution to the entropy, $f(1)=0$.
Passing from (\ref{f}) to (\ref{fn})
is done by adding a (pure topological) GB term to the action  (\ref{action}),
which does not contribute to the equations of motion.
 }
\cite{Wald:1993nt}
\begin{equation}
\label{S}
S=S_{\rm BH}+S_{sGB}\ , \qquad {\rm with}~~~S_E=\frac{A_H}{4},~~~ S_{sGB}=-2\alpha \int_{H} d^2 x \sqrt{h}f(\phi) {\rm  R}^{(2)} \ .
\end{equation} 
In the above relation ${\rm  R}^{(2)}$ is the Ricci scalar of the metric $h_{ij}$, induced on the spatial sections of the event horizon, $H$. 
For the employed ansatz, this results in
\begin{eqnarray}
S=\pi r_H^2  - 16 \alpha \pi  \left( \phi_H \left[1-\frac{\phi^2_H}{3}\right]-\frac{2}{3} \right) \ .
\end{eqnarray}
On the other hand, the ADM mass $M$ 
and the scalar 
` charge' $Q_s$
are determined by the far field asymptotics
 (\ref{inf}).

Finally, let us note that
the equations of the model are invariant under the transformation
\begin{eqnarray}
\label{scale}
r\to \lambda r,~~~\alpha  \to \alpha \lambda^2 \ ,
\end{eqnarray} 
(with $\lambda>0$ an arbitrary constant)
such that only quantities invariant under (\ref{scale})
have a physical meaning.
For example, one can work with the  dimensionless quantities  
\begin{eqnarray}
\label{quant}
 a_H=\frac{A_H}{16\pi M^2},~~t_H=8\pi T_H M ~,
\end{eqnarray} 
such that $a_H=t_H=1$
in the GR limit.

\subsubsection{Generic solutions}

The solutions of the model fall into two different classes,
depending on the asymptotic value   $\phi_\infty$ of the scalar field being unity or not.
In the  generic case, the scalar field does $not$
approach asymptotically the ground state,
 $\phi_\infty \neq  1$. 

Let us remark that
for a small enough scalar field,
 one can construct a perturbative solution  
 as a power series in the dimensionless parameter
\begin{equation}
 \beta =\frac{\alpha}{r_H^2}.
\end{equation}
Assuming 
that the horizon is still located at $r=r_H$,
 one considers
a generic expansion
\begin{equation}
\label{pert}
N (r)=\left(1-\frac{r_H}{r }\right)\sum_{k\geq 0}  \beta^k n_k(r)\ , \qquad 
\delta(r)=\sum_{k\geq  0} \beta^k  \delta_k(r) \ ,
\qquad 
\phi(r)=\sum_{k\geq  1} \beta^k \phi_k(r),
\end{equation}
 the field equations being solved
 order by order in $\beta$.
The functions
$\{n_k (r)$, 
$\delta_k(r)$ 
and  $\phi_k (r)\}$ 
are polynomials in $x=r_H/r$.
The first  few  terms are simple enough, with
 \begin{eqnarray}
\nonumber
&&
n_0(r)=1,~~
n_1(r)=0,~~
n_2(r)=
-\frac{196}{5 }x
-\frac{116}{5 }x^2
-\frac{76}{5 }x^3
+\frac{8123}{15 }x^4
+\frac{872}{15 }x^5
+\frac{184 }{3 }x^6,
\\
\label{perxt}
&&
\delta_1(r)=0,~~
\delta_2(r)=
8{x^2}
 +\frac{32 x^{3}}{3 }
+28 x^{4}
 +\frac{128 x^{5} }{ 5}
+24 x^{6} ,
\\
\nonumber
&&
\phi_1(r)=\phi_{0}-4x -2x^2-\frac{4x^3}{3 },~~\phi_2(r)=0,
\end{eqnarray}
while the $n.l.o.$ terms are too complicated to include here.

 \begin{figure}[h!]
\begin{center}
\includegraphics[width=0.45\textwidth]{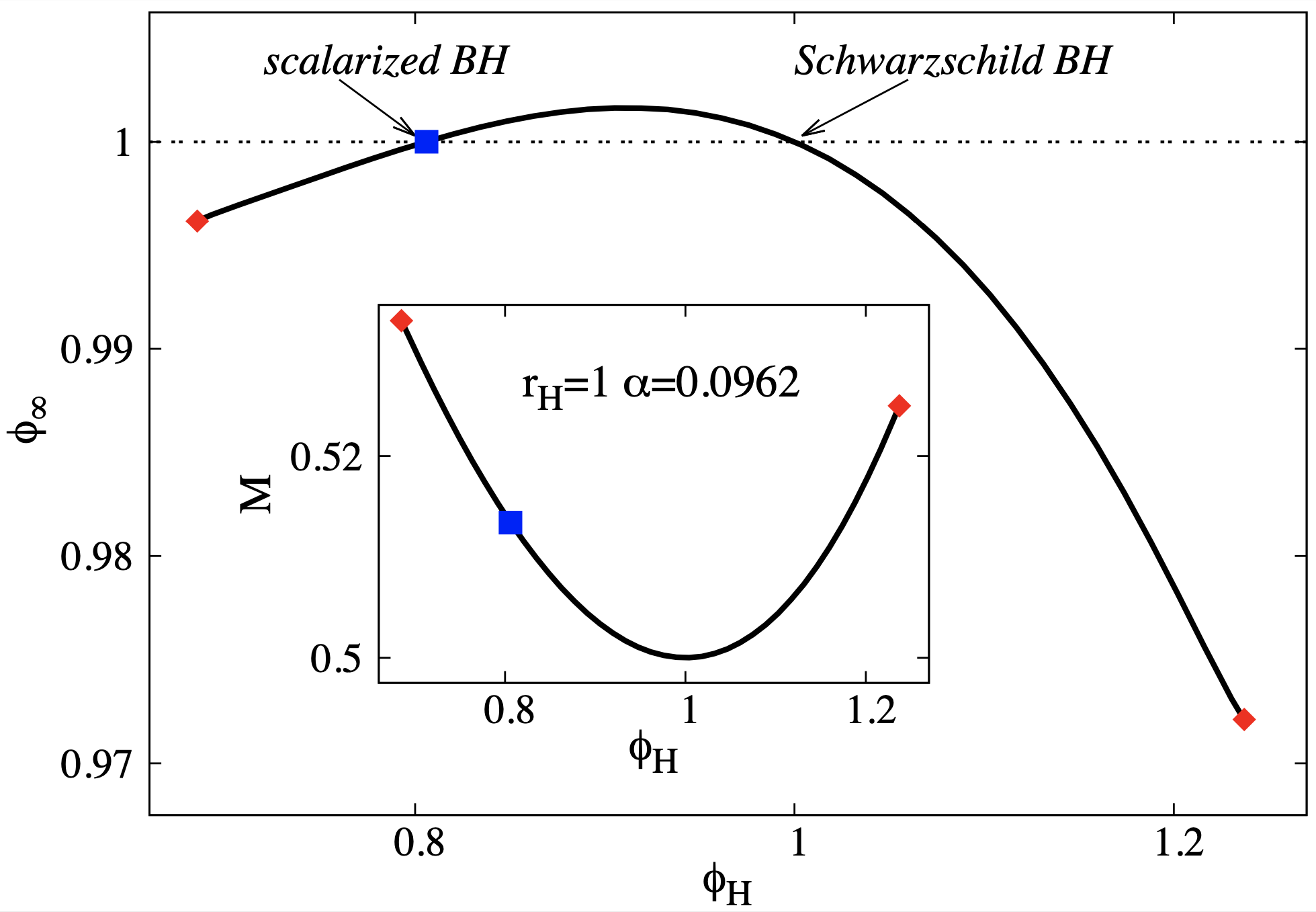}  
\ \ \
\includegraphics[width=0.45\textwidth]{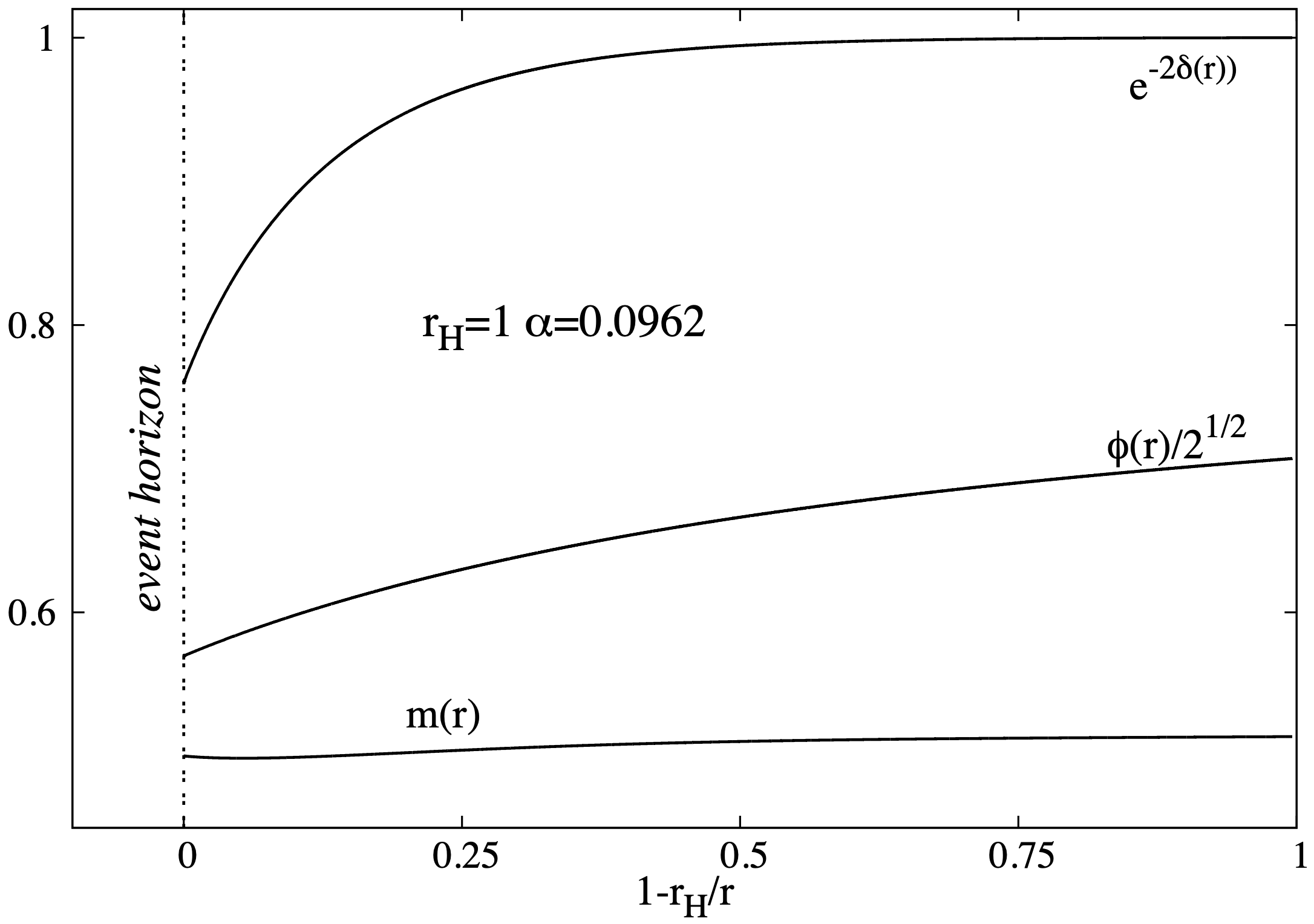} 
\caption{ 
{\it Left panel:} The asymptotic value of the scalar field
$\phi_\infty$
 and the mass $M$
of the solutions are shown as a function of the value of the scalar field 
at the horizon $\phi_H$.
The solutions have a fixed value of the horizon radius
$r_H$
 and of the coupling constant $\alpha$.
{\it Right panel:} The profile of a typical scalarized solution 
(marked with a blue square in the left panel and Figure 2) is shown as a function 
of the compactified coordinate $1-r_H/r$.
}
\label{fig1}
\end{center}
\end{figure} 

We also display the expression of the first few terms for 
several quantities of interest  
\begin{eqnarray}
&&
M=M^{(0)}
\left(
1+
\frac{196 }{5}\beta^2
+\frac{8 P_1(\phi_0)}{3274425}\beta^4 
\right)+\dots,~~
T_H=T_H^{(0)}
\left(
1
- \frac{724}{15  }\beta^2
- \frac{8P_2(\phi_0)}{3274425 }\beta^4
\right)+\dots,
\\
\nonumber
&&
S=S^{(0)}
\left(
1
+ (\frac{352}{3}-16 \phi_0)\beta^2
+P_3(\phi_0) \beta^4 
\right)+\dots,
~~
 ~Q_s=r_H
\left(
4\beta
+
\left(
-\frac{101096}{945}
-\frac{584 \phi_0}{15}
-4\phi_0^2
\right)
\beta^3 
\right)+\dots~,
\end{eqnarray}
with 
\begin{eqnarray}
&&
P_1(\phi_0)=-418524536 + 693 (530552 - 46305 \phi_0) \phi_0,~~
\\
\nonumber
&&
P_2(\phi_0)= 407910422 + 693 \phi_0 (-1693312 + 113715 \phi_0),~~
\\
\nonumber
&&
P_3(\phi_0)= -\frac{143915024}{59535}+
 \frac{16}{225} \phi_0 \left(31222 + 75 (-44 + \phi_0) \phi_0 \right)~,
\end{eqnarray}
and
\begin{equation}
M^{(0)}=\frac{r_H}{2}, ~~
S^{(0)}=\pi r_H^2,~~
T_H^{(0)}=\frac{1}{4\pi r_H}~.
\end{equation}
One can verify  that this solution provides a reasonable approximation for small
$\beta$
and  
$\phi_\infty=\beta \phi_0$.

\medskip

Of special interest, however, are the configurations with large values of $\phi_\infty$,
which are found numerically, 
and may form  
a disconnected branch from that described by the perturbative solution above.
In their construction,
one starts from the expansion (\ref{eh}) and  integrate
towards  $r\to\infty$  the system 
of three EGBs equations by
 using a standard ordinary  
differential  equation solver.
In practice, we integrate up to $r_{\rm max} \simeq 3\times 10^{3}r_H$,
such that  the asymptotic limit 
(\ref{inf}) is reached with enough accuracy.
 Given 
$(r_H,~\phi_\infty;~ \alpha)$, solutions with the right asymptotics may exist for discrete set of  the 
shooting parameters  $(\phi_H,~\delta_H)$, indexed by the node number of the scalar field.
Here we shall report the results for most relevant case of  fundamental, nodeless solutions.

Several results in this case are displayed in Figure \ref{fig1} (left panel),
where we show 
the asymptotic value of the scalar field
$\phi_\infty$ and mass $M$ of the solutions
as a function of the scalar field value at the horizon, $\phi_H$.
One can see that the solutions exist for a finite range of $\phi_H$
(located around the ground state $\phi\equiv 1$),
ending in critical configurations where the condition (\ref{cond})
fails to be satisfied\footnote{This last feature cannot be capture within a perturbative approach.}.

 \begin{figure}[h!]
\begin{center}
\includegraphics[width=0.45\textwidth]{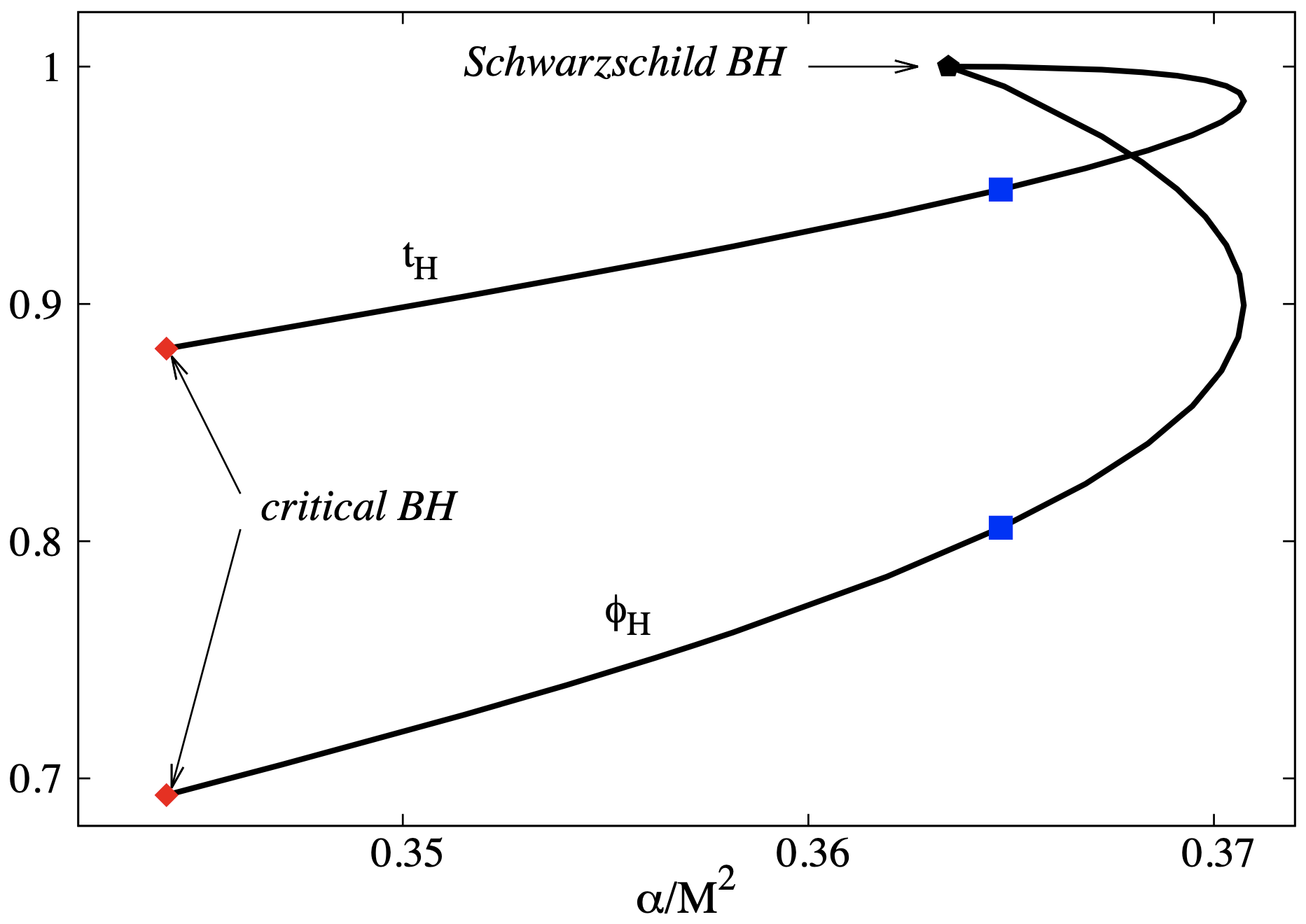}  
\ \ \
\includegraphics[width=0.45\textwidth]{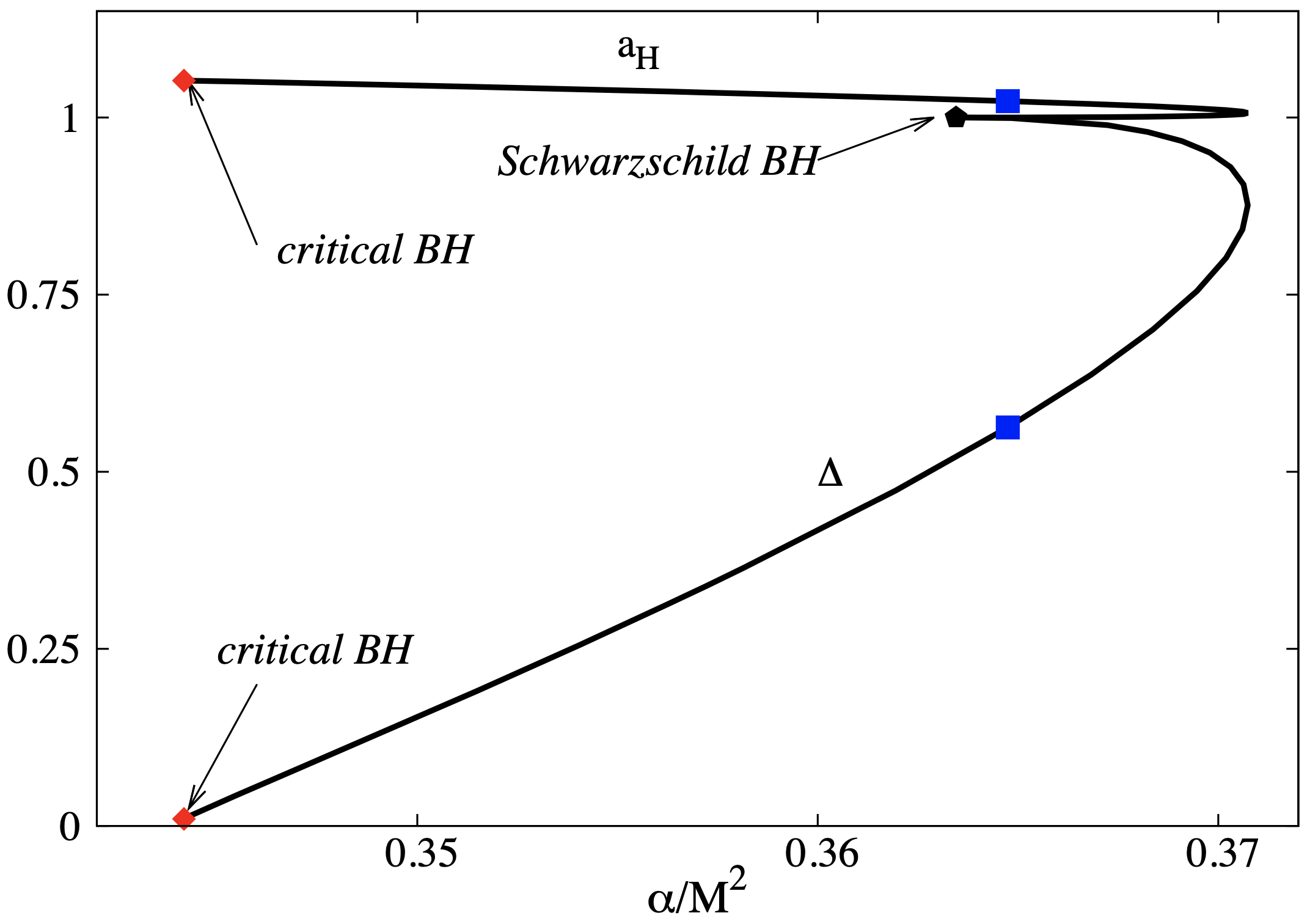} 
\caption{ 
Several quantities of interest are shown as a function of the ratio
$\alpha/M^2$ for the set of scalarized BHs.
}
\label{fig2}
\end{center}
\end{figure} 

\subsubsection{Scalarized BHs}

As one can see in Figure
\ref{fig1} (left panel),
a particular configuration there has
 $\phi_{\infty}=1$,
while $\phi_H\neq 1$.
This configuration is of special interest, corresponding to a scalarized BH,
its profile being displayed in Figure \ref{fig1} (right panel).

That solution, however,  has no special features;
similar configurations are found for a range of the   parameters $\alpha,r_H$.
Also,   as with the generic case, only nodeless solutions were studied so far, 
corresponding to the fundamental states; 
however, solutions with nodes exist as well, corresponding to excited EGBs configurations.

The basic properties of the  scalarized BHs are rather similar to those 
found in \cite{Silva:2017uqg} for a quadratic coupling function
They can
be summarized as follows (see Figure \ref{fig2}). 
First,
these spherically symmetric BHs bifurcate from a special Schwarzschild solution 
supporting the scalar cloud 
($i.e.$ an infinitesimally small scalar field), 
with $\alpha/M^2 \simeq 0.3635$.
Second, keeping constant the parameter $\alpha$ (or, equally, the event horizon radius $r_H$),
the solutions can be obtained continuously in the parameter space,
forming a line, which starts from the smooth GR Schwarzschild limit ($\phi \equiv 1$), and ends at a limiting solution.
Once the limiting configuration is reached, the solutions cease to exist in
the parameter space.
The existence of the  `critical'
configuration   can be understood from the condition
(\ref{cond}), with the determinant $\Delta$ vanishing at that point.
Finally, as with the solutions in Ref. \cite{Silva:2017uqg},
for a given ADM mass, the entropy of the solutions
is maximized by the Schwarzschild vacuum BH
 (although the relative difference is rather small, of order $10^{-3}$).

\subsection{The scalar-vector model: perturbative solutions}

The full model (\ref{action})
contains also a vector field $A$,
which has been consistently set to zero in the above study.
The investigation of self-gravitating   configurations
with $A\neq 0$ is a complicated task, which we do not attempt to
address here.
Instead, we report the results for a preliminary investigation of
 scalar-vector solutions in  the `probe limit',
$i.e.$  is for a fixed Schwarzschild BH background,
as given by $\delta=0$, $N(r)=1-r_H/r$ in the metric Ansatz (\ref{ansatz}).
This case is technically much simpler, while the corresponding solutions
are likely to
capture some of the basic properties of the full model.

 \begin{figure}[h!]
\begin{center}
\includegraphics[width=0.45\textwidth]{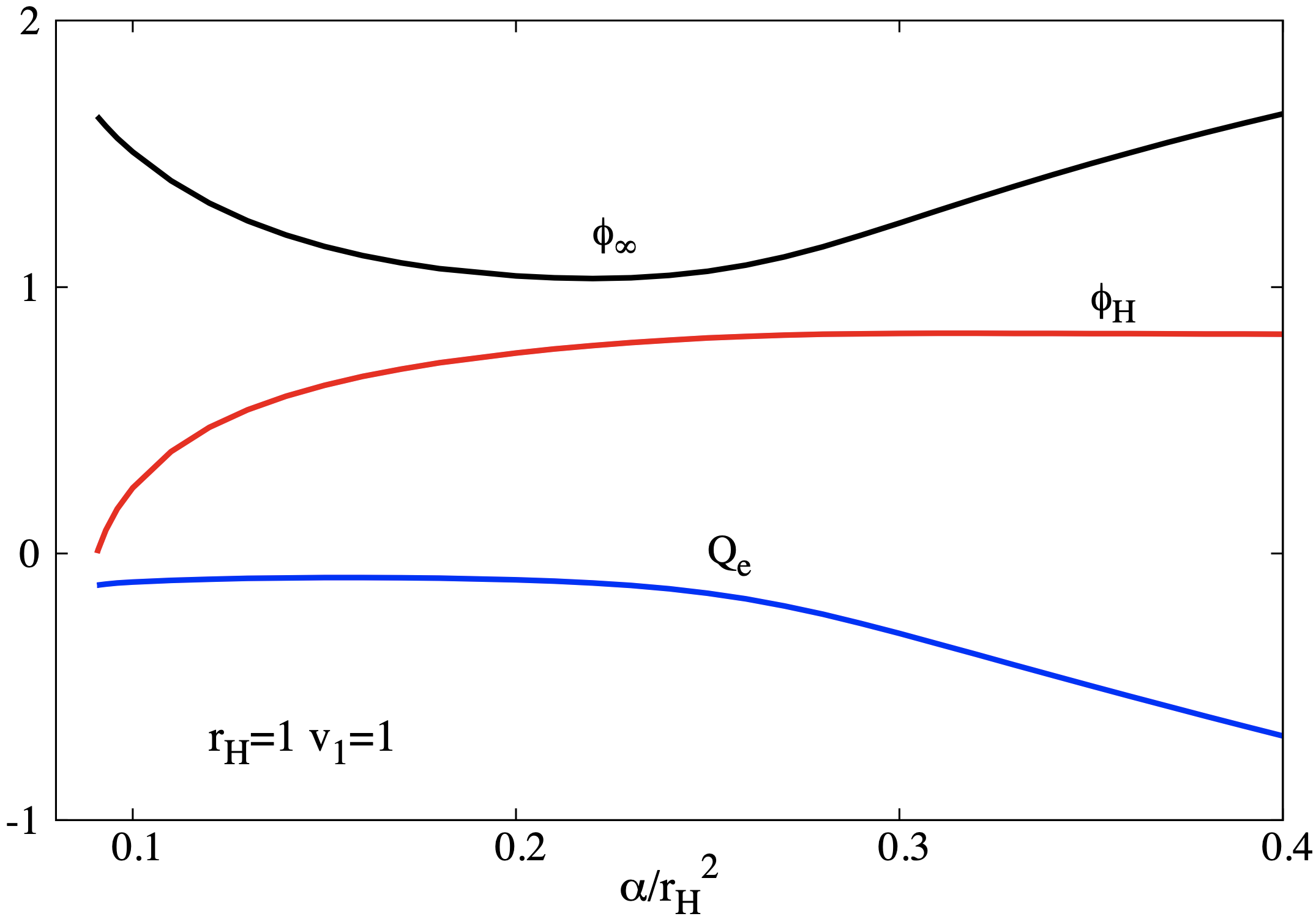}  
\ \ \
\includegraphics[width=0.45\textwidth]{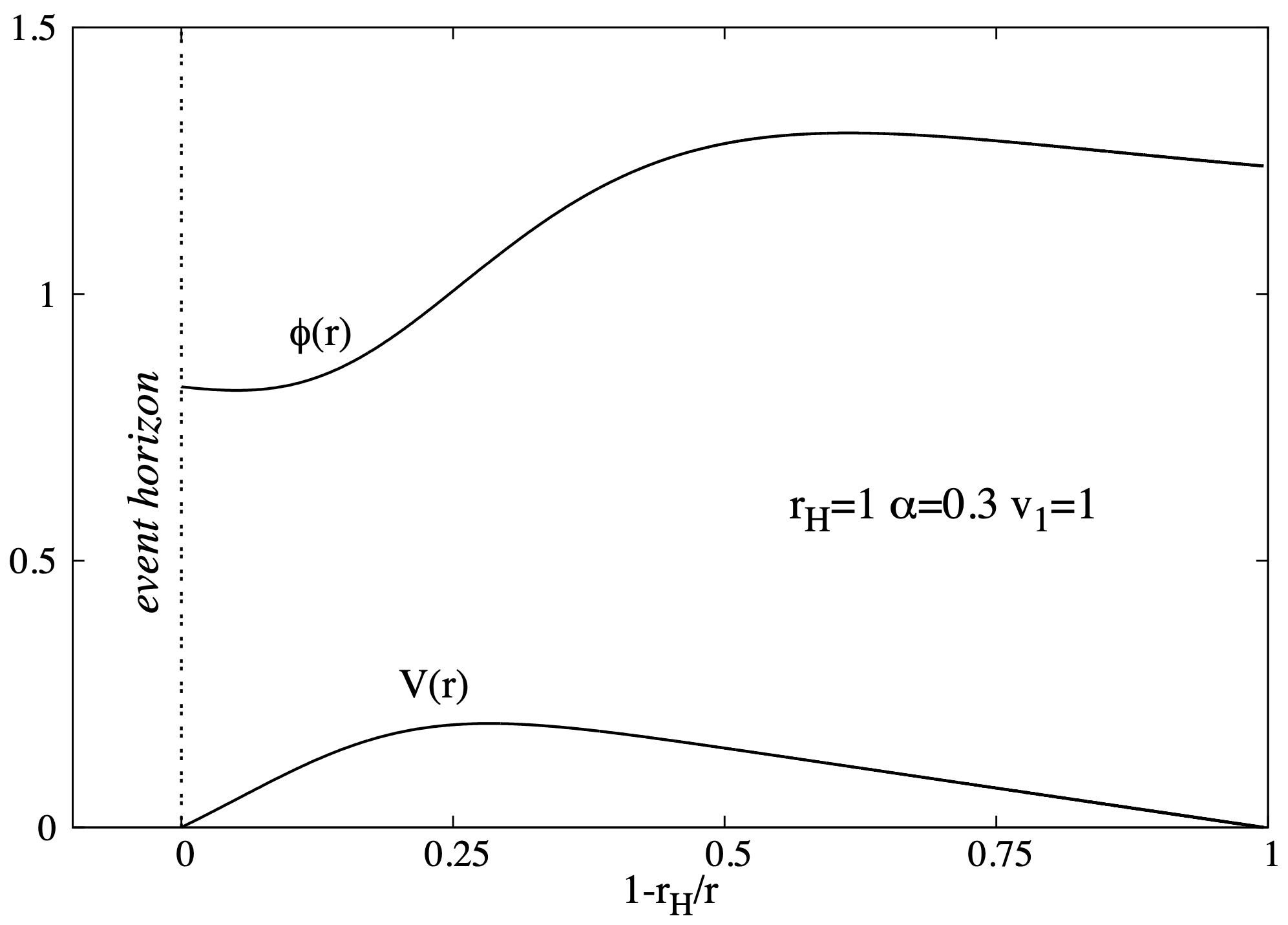} 
\caption{ 
{\it Left panel:} The asymptotic value of the scalar field
$\phi_\infty$, the value of the scalar field at the horizon $\phi_H$
and the charge of the vector field $Q_e$
  are shown as a function of the  ration $\alpha/r_H^2$
	for solutions of the scalar-vector model in the probe limit.
	{\it Right panel:}
	The profile of a typical solution
	of the scalar-vector model
	in a fixed Schwarzschild BH background  
is shown as a function 
of the compactified coordinate $1-r_H/r$.
For all solutions  the vector field vanishes both at the horizon 
and at infinity. 
}
\label{fig3}
\end{center}
\end{figure} 

Restricting again to the spherically symmetric case, we consider an ansatz with
 \begin{eqnarray}
\phi \equiv \phi(r),~~ A=V(r) dt,
\end{eqnarray}
which can be shown to be consistent.
Then, when ignoring the backreaction,  
the problem reduces to solving
the following coupled equations
 \begin{eqnarray}
\label{eqs-test}
&&
\phi''+\frac{2r-r_H}{r(r-r_H)}\phi'
+\frac{16 \alpha r_H^2}{r^5(r-r_H)}
\left(
3(\phi^2-1)
-\frac{r(3r-2r_H)V^2}{(r-r_H)^2}
+\frac{2r^2 V V'}{(r-r_H)}
\right)=0,
\\
\nonumber
&&
V''+\frac{2V'}{r}
+\frac{32 \alpha r_H^2 \phi'V}{r^4(r-r_H)}=0,
\end{eqnarray}
The approximate form of the solutions close to the horizon reads
 \begin{eqnarray}
\label{nh}
&&
 \phi(r)=\phi_H -\frac{16 \alpha (r_H^2 v_1^2+3(\phi_H^2-1))}{r_H^3}(r-r_H)+O(r-r_H)^2,
\\
\nonumber
&&
V(r)=v_1(r-r_H)-\frac{v_1 }{r_H}
\big(1-\frac{256 \alpha^2}{r_H^4} (r_H^2 v_1^2+3(\phi_H^2-1)) \big)
(r-r_H)^2+O(r-r_H)^3,
\end{eqnarray}
where $\phi_H$ and $v_1$ are arbitrary constants.
For large-$r$, the first terms in a $1/r$ expansion of the solution reads 
 \begin{eqnarray}
 \phi(r)=\phi_\infty+\frac{Q_s}{r}+O(1/r^2),
~~
V(r)=V_\infty+\frac{Q_e}{r}+O(1/r^5),
\end{eqnarray}
with $\phi_\infty$,
$V_\infty$,
$Q_e$ and $Q_s$
free parameters.

Eqs. (\ref{eqs-test})
are solved numerically, by using the same approach as in the EGBs case.
The numerical results
 indicate the existence, for any $\alpha$, of a continuum of solutions in terms of
the constants $\phi_H$, $v_1$
which enter the near horizon expansion (\ref{nh}).
The asymptotic values of the scalar and vector fields depend of the  input parameters
 $\phi_H$, $v_1$, $r_H$ and $\alpha$
(note that, as expected from the structure of the equations,
the vector field vanishes identically, 
when taking $v_1=0$).

Of special interest here are the configurations
which approach asymptotically the vector ground state,
$i.e.$ 
 with $V_\infty=0$.
In Figure \ref{fig3} (left panel) we display
some quantities of interest for a set of such solutions;
there the 
values of $r_H$
and $v_1$ are fixed,
while the parameter
$\alpha$
is varied
(note that we restrict again to the case of nodeless configurations).
One can notice that, for a given $\alpha$,
the solution with $V(\infty)=0$
is found for a single value of $\phi_H$
while $\phi_\infty \neq 1$.
The profile of a typical solution is shown in  Figure \ref{fig3} (right panel).

The presence of  probe-limit solutions with $V_\infty=0$ 
 provides a  hint for the existence of vectorized configurations in the full 
backreacting model.
That is,
the  generic solutions above with $\phi_\infty \neq 1$
may possess generalizations with a nonzero vector field
in the bulk, and still approach asymptotically
the vector ground state\footnote{See thee recent work \cite{Barton:2021wfj}
for a study of vectorized BHs together with a review of the literature.}.
We hope to return to the study of these configurations elsewhere.

\section{Further remarks}
The main purpose of this work was to point out that families of EGBs models with coupling functions  that  permit spontaneous scalarization  
can be  motivated by a geometric/topological construction~\cite{Tchrakian:2017fdw,Radu:2020ytf}. This  indicates a more natural embedding  for GB spontaneous scalarization models; typically the  necessary
coupling is simply postulated \textit{ad hoc}. Moreover, as a case study, we present a preliminary investigation 
of the scalarized BH solutions in a particular EGBs model emerging from this sort of construction. 

The models discussed herein offer both  novel features and challenges. As for the novel features, the corresponding coupling function
is a sum of a linear and a cubic term,
with the scalar field possessing two disconnected ground states.
A consequence of this fact is that the BH 
scalarization becomes possible for any sign of the 
coupling constant $\alpha$
in front of the GB term (but around a different vacuum for each sign). As for the challenges, the cubic term (or, in general, the highest odd power term) raises the issue of the stability of  the model. Moroever, the construction is  not  complete: it provides the GB term  and the non-minimal couplings, but the resulting HCSG term has been added \textit{ad hoc}
to the standard Einstein-scalar field action.

The basic properties of the scalarized BHs constructed herein
were found to be rather similar to 
those in the original work \cite{Silva:2017uqg},
where a quadratic coupling function has been employed.
In particular, the scalarized BHs are entropically
disfavoured  over the Schwarzschild vacuum configuration.
Generic solutions (with a scalar field which does not 
approach asymptotically the ground state)
were also studied, and a perturbative solution was reported.

A preliminary investigation of the solutions of the general model 
in   \cite{Tchrakian:2017fdw}
	with an extra-vector field 
	has been also considered.
Only  the
 probe limit regime of the solutions ($i.e.$ for a fixed Schwarzschild background)
has been considered in this case, 
the  results hinting towards the possible existence of 
'vectorized' BHs within the full model.

\bigskip
\bigskip

\noindent {\large\bf Acknowledgements}
\\  
The  work of E.R.  is  supported  by  the Center  for  Research  and  Development  in  Mathematics  and  Applications  (CIDMA)  
through  the Portuguese Foundation for Science and Technology (FCT - Fundacao para a Ci\^encia e a Tecnologia), 
references UIDB/04106/2020 and UIDP/04106/2020, and by national funds (OE), through FCT, I.P., 
in the scope of the framework contract foreseen in the numbers 4, 5 and 6 of the article 23,of the Decree-Law 57/2016, of August 29, changed by Law 57/2017, of July 19.  
We acknowledge support  from  the  projects  PTDC/FIS-OUT/28407/2017,  CERN/FIS-PAR/0027/2019 and PTDC/FIS-AST/3041/2020.  
 This work has further been supported by the European Union’s Horizon 2020 research and innovation (RISE) programme H2020-MSCA-RISE-2017 Grant No. FunFiCO-777740.  
The authors would like to acknowledge networking support by the COST Action CA16104.

\begin{small}

\end{small}

\end{document}